\documentclass[superscriptaddress,aps,pra,twocolumn,showpacs,preprintnumbers,amsmath,amssymb,floatfix,groupedaddress]{revtex4}

\usepackage[dvips]{graphicx}
\usepackage{dcolumn}
\usepackage{bm}
\usepackage{amssymb}
\usepackage{xr}

\newcommand{\eop}{\mathcal{E}}

\begin{document}

\title{Photon storage in $\Lambda$-type optically dense atomic media. III. Effects of inhomogeneous broadening}

\author{Alexey V. Gorshkov}
\author{Axel Andr\'e}
\author{Mikhail D. Lukin}
\affiliation{Physics Department, Harvard University, Cambridge, Massachusetts 02138, USA}
\author{Anders S. S{\o}rensen}
\affiliation{QUANTOP, Danish National Research Foundation Centre of
Quantum Optics, Niels Bohr Institute, DK-2100 Copenhagen {\O},
Denmark}

\date{\today}


\begin{abstract}
In a recent paper [Gorshkov \textit{et al.}, Phys. Rev. Lett. \textbf{98}, 123601 (2007)] and in the two preceding papers [Gorshkov \textit{et al.}, Phys. Rev. A \textbf{76}, 033804 (2007); \textbf{76}, 033805 (2007)], we used a universal physical picture to optimize and demonstrate equivalence between a wide range of techniques for storage and retrieval of photon wave packets in homogeneously broadened $\Lambda$-type atomic media, including the adiabatic reduction of the photon group velocity, pulse-propagation control via off-resonant Raman techniques, and photon-echo-based techniques. In the present paper, we generalize this treatment to include inhomogeneous broadening. In particular, we consider the case of Doppler-broadened atoms and assume that there is a negligible difference between the Doppler shifts of the two optical transitions. In this situation, we show that, at high enough optical depth, all atoms contribute coherently to the storage process as if the medium were homogeneously broadened. We also discuss the effects of inhomogeneous broadening in solid state samples. In this context, we discuss the advantages and limitations of reversing the inhomogeneous broadening during the storage time, as well as suggest a way for achieving high efficiencies with a nonreversible inhomogeneous profile.
\end{abstract} 



\maketitle

\section{Introduction}

The faithful storage of a traveling light pulse in an atomic memory and the subsequent retrieval of the state are currently being pursued in a number of laboratories around the world. A central question that emerges from these studies is which approach represents the best possible strategy and how the maximum efficiency can be achieved. 
In a recent paper \cite{gorshkov07}, we used a novel universal physical picture to optimize and demonstrate equivalence between a wide range of techniques for storage and retrieval of photon wave packets in $\Lambda$-type atomic media in free space, including the adiabatic reduction of the photon group velocity, pulse-propagation control via off-resonant Raman techniques, and photon-echo-based techniques. In two preceding papers Refs.~\cite{paperI,paperII}, which we will refer to henceforth as papers I and II, we present
the full details of the analysis of Ref.~\cite{gorshkov07} as well as several of its extensions,
both for an ensemble enclosed in a cavity and for the free-space model. While the analysis of papers I and II is limited to homogeneously broadened media, many experimental realizations, such as the ones using warm atomic vapors \cite{eisaman05} or the ones using impurities in solid state samples \cite{manson06,afzelius06}, have some degree of inhomogeneous broadening, whose presence will modify the optimal control strategy and the values for the maximum efficiency. The subject of the present paper is the extension of the analysis of papers I and II to include inhomogeneous broadening.

The remainder of the present paper is organized as follows. In Sec.~\ref{sec:Doppler}, we discuss the effects of inhomogeneous broadening assuming that the atoms fully redistribute themselves between frequency classes during the storage time, which would be the case, for example, in Doppler-broadened atoms with sufficiently long storage times. In particular, we optimize the storage process and show that at high enough optical depth, all atoms contribute coherently as if the medium were homogeneously broadened. Then in Sec.~\ref{sec:inhom}, we discuss the effects of inhomogeneous broadening assuming there is no redistribution between frequency classes during the storage time, which would be the case in atomic vapors for short storage times or in solid state samples. In particular, we discuss the advantages and limitations of reversing the inhomogeneous broadening during the storage time \cite{kraus06}, as well as suggest a method for achieving high efficiencies with a nonreversible spectrally well-localized inhomogeneous profile. In Sec.~\ref{sec:inhomsum}, we summarize our analysis of the effects of inhomogeneous broadening. Finally, in the Appendix, we present some details omitted in the main text.

\section{Inhomogeneous Broadening with Redistribution between Frequency Classes during the Storage Time \label{sec:Doppler}}

In this section, we consider a particular case of inhomogeneously broadened media: the case of a Doppler-broadened atomic vapor in free space. We first describe our model in Sec.~\ref{sec:DopplerModel}. We then use this model in Sec.~\ref{sec:dopretst} to analyze storage and retrieval of photons in Doppler-broadened media. 

\subsection{Model \label{sec:DopplerModel}}

\begin{figure}[tb]
\includegraphics[scale = 0.5]{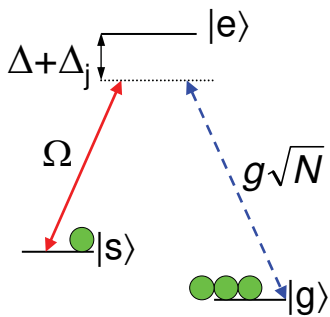}
\caption{(Color online) $\Lambda$-type medium coupled to a classical field (solid) with Rabi frequency $\Omega(t)$ and a quantum field (dashed). Collective enhancement \cite{lukin03} effectively increases the atom-field coupling constant $g$ up to $g \sqrt{N}$, where $N$ is the number of atoms in the medium. $\Delta_j$ is a frequency shift due to inhomogeneous broadening. \label{fig:lambda}}
\end{figure}

As in paper II, we consider a free-space medium of length $L$ and cross-section area $A$ containing $N = \int_0^L d z n(z)$ atoms, where $n(z)$ is the number of atoms per unit length. We assume that within the interaction volume the concentration of atoms is uniform in the transverse direction. The atoms have the $\Lambda$-type level configuration shown in Fig.~\ref{fig:lambda}. They are coupled with a collectively enhanced coupling constant $g \sqrt{N}$  ($g$ assumed to be real for simplicity) to a narrowband quantum field centered at a frequency $\omega_1$ and described by a slowly varying operator $\hat \eop(z,t)$. The atoms are also coupled to a copropagating narrowband classical control field centered at frequency $\omega_2$ with a Rabi frequency envelope $\Omega(z,t) = \Omega(t-z/c)$. We assume that quantum electromagnetic field modes with a single transverse profile are excited. As discussed in detail in paper II, we neglect reabsorption of spontaneously emitted photons and treat the problem in a one-dimensional approximation.    

In order to model the inhomogeneous broadening, we sort all the atoms into separate velocity classes labeled by $j$, such that all atoms in a certain velocity class have approximately the same velocity $v_j$. We will assume that during the experiment the atoms do not change their positions significantly compared to the longitudinal variation of the fields. Therefore, the Doppler effect will be the only effect of nonzero atomic velocities that we will consider.  
 
Furthermore, we will assume that the difference in the Doppler shifts of the two optical transitions can be neglected so that atoms in all velocity classes can simultaneously stay in two-photon resonance with the two fields. For copropagating beams, this assumption is exactly satisfied if the two metastable states are degenerate. Moreover, below we shall consider a storage technique which we refer to as fast storage \cite{gorshkov07, paperII}, where the control field consists of a simple $\pi$ pulse, which will work perfectly provided its Rabi frequency is much greater than the broadened linewidth.  In this case, the assumption of equal Doppler shifts of the two optical transitions is thus not needed. However, we shall also consider the so-called adiabatic storage schemes \cite{gorshkov07, paperII}, where the splitting of the two metastable levels by a nonzero frequency difference $\omega_{sg}$, as well as Doppler broadening occurring for noncopropagating fields, will play a role \cite{shuker07}. In order to ensure that the difference in Doppler shifts has a negligible effect in this situation, we consider copropagating beams and assume that the total accumulated phase difference $T \overline{v} \omega_{sg}/c$ is much less than unity, where $\overline{v}$ is the  thermal speed of the atoms, $c$ is the speed of light, and $T$ is the duration of the incoming quantum light pulse. This condition is usually satisfied even in room-temperature experiments, such as the experiment using $^{87}$Rb vapor in Ref.~\cite{eisaman05}, where $\omega_{sg} = (2 \pi) 6.8 \textrm{ GHz}$ and $T \sim 200 \, \mu\textrm{s}$, which gives $T \overline{v} \omega_{sg}/c < 0.01$. 

For simplicity, we also assume that the velocities do not change during the processes of storage and retrieval but fully rethermalize during the storage time $[T,T_\textrm{r}]$. The atoms thus fully redistribute themselves among different velocity classes during the storage time, and at time $T_\textrm{r}$ the spin wave is the same across all velocity classes. In addition to being relevant for Doppler broadening, much of the discussion of this section will also apply to solid state systems with inhomogeneous broadening. The fact that solid state impurities do not have the redistribution between frequency classes during the storage time (which we assume in this section) will, however, introduce modifications. Some of these modification are discussed in Sec.~\ref{sec:inhom}. 

We assume that the control is detuned by $\Delta$ with respect to stationary atoms, $\omega_2 = \omega_{es}-\Delta$, while the quantum field is in two-photon resonance, i.e., $\omega_1 = \omega_{eg}-\Delta$ (where $\omega_{es}$ and $\omega_{eg}$ are atomic transition frequencies). We define the same slowly varying operators as in Eqs.~(A7)-(A11) in paper II, except now, at each $z$ and $t$, we have continuous atomic operators for each velocity class. For example, we define
\begin{equation}
\hat \sigma_{\mu\mu}^j(z,t) = \sqrt{p_j} \frac{1}{N_z^j} \sum_{i=1}^{N_z^j} \hat \sigma^{j (i)}_{\mu\mu}(t),
\end{equation}
where $j$ indicates the velocity class, $i$ runs over the $N_z^j$ atoms near $z$ in the velocity class $j$, $p_j$ is the fraction of atoms in the velocity class $j$, $\sum_j p_j = 1$, and $\sigma^{j (i)}_{\mu\nu} = |\mu\rangle_{j (i)} \langle \nu|$ indicates the internal state operator between states $|\mu\rangle$ and $|\nu\rangle$ for the $i$th atom in the $j$th velocity class.

As in papers I and II, we use the dipole and rotating-wave approximations, assume that almost all atoms are in the ground state at all times, and consider equations of motion only to first order in $\hat \eop$. As described in papers I and II, under reasonable experimental conditions, the incoming noise corresponding to the decay of atomic coherences is vacuum, and efficiency is the only number we need in order to fully characterize the mapping. Following then the same steps as in Sec.~II of paper II, we obtain complex number equations 
\begin{eqnarray} \label{doppler1eqe}
\!\!\!\!\!\!\!(\partial_t + c \partial_z) \eop \!\!&=&\!\! i g \sqrt{N} \sum_j \sqrt{p_j} P_j n(z) L/N,
\\ \label{doppler1eqp}
\partial_t P_j \!\!&=&\!\! - (\gamma\!+ \!i (\Delta\!+\!\Delta_j)) P_j \!+\! i g \sqrt{N}\! \sqrt{p_j} \eop \!+\! i \Omega S_j,
\\ \label{doppler1eqs}
\partial_t S_j \!\!&=&\!\! i \Omega^* P_j,
\end{eqnarray}
where $\gamma$ is the polarization decay rate, $\Delta_j = \omega^j_{es}-\omega_{es} = \omega_{es} v_j/c$ is the Doppler shift due to the velocity $v_j$ of the $j$th velocity class, $P_j$ is the complex number representing the optical polarization $\sqrt{N} \hat \sigma^j_{ge}$, and $S_j$ is the complex number representing the spin wave $\sqrt{N} \hat \sigma^j_{gs}$. We assume in Eq.~(\ref{doppler1eqs}) that the decay rate of the spin wave $S_j$ is negligible. As in Eqs.~(5)-(7) in paper II, we now go into the comoving frame $t' = t - z/c$, introduce the dimensionless time $\tilde t = \gamma t'$ and a dimensionless rescaled coordinate $\tilde z = \int_0^z d z' n(z')/N$, absorb a factor of $\sqrt{c/(L \gamma)}$ into the definition of $\eop$, and obtain
\begin{eqnarray} \label{doppler2eqe}
\partial_{\tilde z} \eop &=& i \sqrt{d} \sum_j \sqrt{p_j} P_j,
\\ \label{doppler2eqp}
\partial_{\tilde t} P_j &=& - (1+ i(\tilde \Delta + \tilde \Delta_j)) P_j + i \sqrt{d} \sqrt{p_j} \eop + i \tilde \Omega S_j,
\\ \label{doppler2eqs}
\partial_{\tilde t} S_j &=& i \tilde \Omega^* P_j,
\end{eqnarray}
where we have identified the optical depth $d = g^2 N L/(\gamma c)$, and where $\tilde \Omega = \Omega/\gamma$, $\tilde \Delta = \Delta/\gamma$, and $\tilde \Delta_j = \Delta_j/\gamma$. Note that $d$ is defined here as the optical depth that the sample would have had if there were the same number of atoms but no inhomogeneous broadening. This quantity should not be confused with the actually measured optical depth $d'$. With inhomogeneous broadening, the measured value will be lower: $d'<d$. Later, we shall derive explicit relations between these two quantities.  It is essential to realize that both quantities $d$ and $d'$ play a role in the performance of the memory, as we shall see below.  

We assume that all atoms are initially pumped into the ground state, i.e., no $P$ or $S$ excitations are present in the atoms. We also assume that the only input field excitations initially present are in the quantum field mode with a normalized envelope shape $\eop_\textrm{in}(\tilde t)$ nonzero on $[0,\tilde T]$ (where $\tilde T = T \gamma$). The goal is to store the state of this mode into some spin-wave mode and at a time $\tilde T_\textrm{r} > \tilde T$ retrieve it onto an output field mode. The initial conditions for storage are $S_j(\tilde z,0) = 0$ and $P_j(\tilde z,0) = 0$ for all $j$ and $\eop(0,\tilde t) = \eop_\textrm{in}(\tilde t)$, and the storage efficiency is given by 
\begin{equation}
\eta_{\textrm{s}} = \frac{(\textrm{number of stored excitations})}{(\textrm{number of incoming photons})} 
= \int_0^1 d \tilde z |S(\tilde z, \tilde T)|^2,
\end{equation}
where $S(\tilde z, \tilde T) = \sum_j \sqrt{p_j} S_j(\tilde z, \tilde T)$ is the spin wave, to which all $S_j$ average after rethermalization. The initial conditions for retrieval are $\eop(0,\tilde t) = 0$, and, for all $j$, $P_j(\tilde z,\tilde T_\textrm{r}) = 0$, and $S_j(\tilde z,\tilde T_\textrm{r}) = \sqrt{p_j} S(\tilde z,\tilde T)$ or $S_j(\tilde z,\tilde T_\textrm{r}) = \sqrt{p_j} S(1-\tilde z,\tilde T)$ for forward or backward retrieval, respectively (as in paper II). The total efficiency of storage followed by retrieval is then given by
\begin{equation} \label{etatot}
\eta_{\textrm{tot}} = \frac{(\textrm{number of retrieved photons})}{(\textrm{number of incoming photons})}=\int_{\tilde T_\textrm{r}}^\infty \! d \tilde t |\eop_\textrm{out}(\tilde t)|^2,
\end{equation}
where $\eop_\textrm{out}(\tilde t) = \eop(1,\tilde t)$. If during retrieval we instead start with $S_j(\tilde z,\tilde T_\textrm{r}) = \sqrt{p_j} S(\tilde z)$ for some normalized spin wave $S(\tilde z)$, then the same equation will give the retrieval efficiency from this mode:
\begin{equation}
\eta_{\textrm{r}} = \frac{(\textrm{number of retrieved photons})}{(\textrm{number of stored excitations})}
\!=\! \int_{\tilde T_\textrm{r}}^\infty d \tilde t |\eop_\textrm{out}(\tilde t)|^2. 
\end{equation}

For completeness, we note that, for the cavity model described in paper I, the equations corresponding to Eqs.~(\ref{doppler2eqe})-(\ref{doppler2eqs}) above are (without rescaling of variables)
\begin{eqnarray} \label{doppler2eqecavity}
\eop_\textrm{out} &=& \eop_\textrm{in} + i \sqrt{2 \gamma C} \sum_j \sqrt{p_j} P_j,
\\
\partial_t P_j &=& - (\gamma + i(\Delta + \Delta_j)) P_j - \gamma C \sqrt{p_j} \sum_k \sqrt{p_k} P_k 
\nonumber \\ \label{doppler2eqpcavity}
&&+ i \Omega S_j + i\sqrt{2 \gamma C} \sqrt{p_j} \eop_\textrm{in},
\\ \label{doppler2eqscavity}
\partial_t S_j &=& i \Omega^* P_j,
\end{eqnarray}
where $C$ is the cooperativity parameter equal to the optical depth of the ensemble multiplied by the cavity finesse. The initial conditions for storage in the cavity model are $S_j(0) = 0$  and $P_j(0) = 0$ for all $j$ and $\eop_\textrm{in}(t) \neq 0$, while the initial conditions for retrieval are $S_j(T_\textrm{r}) = \sqrt{p_j} S(T)$ and $P_j(T_\textrm{r}) = 0$ for all $j$ and $\eop_\textrm{in}(t) = 0$. It is assumed that during the storage time atoms rethermalize and all $S_j$ average to the same value $S(T)=\sum_j \sqrt{p_j} S_j(T)$.

In the homogeneously broadened case discussed in papers I and II, we defined the so-called adiabatic and fast regimes for storage and retrieval. Both of these limits can also be achieved in the presence of Doppler broadening. The adiabatic regime corresponds to smooth control and input fields such that the optical polarization $P_j$ in Eq.~(\ref{doppler2eqp}) (or Eq.~(\ref{doppler2eqpcavity}) for the case of the cavity model) can be adiabatically eliminated. The fast regime corresponds to storage and retrieval techniques in which the control field consists of a very short and powerful resonant $\pi$ pulse between states $|s\rangle$ and $|e\rangle$. The only difference in the requirements from the homogeneously broadened case is that now the control field $\Omega$ in the $\pi$ pulse must also be much greater than the inhomogeneous width. We refer the reader to paper I for a detailed discussion of the adiabatic and fast photon storage techniques, as well as for a full list of references.

Equations (\ref{doppler2eqe})-(\ref{doppler2eqs}) can be solved numerically by introducing sufficiently many discrete velocity classes. However, when the control is constant (i.e., a step) or when we are in the fast limit, these equations can be solved without discretizing the velocity distribution by using Laplace transformation in time, $\tilde t \rightarrow v$. In this case, the inverse Laplace transform has to be taken numerically at the end. Alternatively, in the case of retrieval alone or in the case of storage followed by retrieval, if one is interested only in the efficiency and not in the output mode, one can compute this efficiency both in the free-space model and in the cavity model without computing $\eop_\textrm{out}(\tilde t)$ directly from the Laplace transform $\eop_\textrm{out}(v)$ as
\begin{equation} \label{etalaplace}
\eta_\textrm{r} = \frac{1}{2 \pi} \int_{-\infty}^\infty d \xi \left|\eop_\textrm{out}(v = i \xi)\right|^2.
\end{equation} 
Below we shall use both the numerical method with discrete velocity classes and the method of the Laplace transformation in time.

\subsection{Retrieval and storage with Doppler broadening \label{sec:dopretst}}

An important result in the discussion of homogeneously broadened ensembles in papers I and II was the result that the retrieval efficiency is independent of the shape of the control and the detuning, provided that all excitations are pumped out of the system. Moreover, in both the cavity and the free-space cases we were able to deduce in papers I and II explicit formulas for the retrieval efficiency. Although in the inhomogeneously broadened case discussed in the present paper we have not been able to find an explicit formula for the retrieval efficiency, we will present now the proof that even with inhomogeneous broadening the retrieval efficiency is independent of the detuning $\Delta$ and the control shape $\Omega(t)$.

We consider first the cavity model given in Eqs.~(\ref{doppler2eqecavity})-(\ref{doppler2eqscavity}). Since we are interested in retrieval, we set $\eop_\textrm{in} = 0$. We also consider a general situation, in which $S_j(t=0)$ are not necessarily equal for different velocity classes $j$ (we also shifted for simplicity the beginning of retrieval from $t = T_\textrm{r}$ to $t=0$). Then, using Eq.~(\ref{doppler2eqecavity}), the retrieval efficiency is
\begin{equation} \label{effdopcav}
\eta_\textrm{r} = \int_0^\infty \!\!\!\! d t \left|\eop_\textrm{out}(t)\right|^2 = 2 \gamma C \sum_{j,k} \sqrt{p_j p_k} \int_0^\infty dt P_j(t) P^*_k(t),
\end{equation}
and the following identity can be explicitly verified from Eqs.~(\ref{doppler2eqpcavity}) and (\ref{doppler2eqscavity}):
\begin{eqnarray} \label{pjpk}
\frac{d}{dt} (P_j P^*_k + S_j S^*_k) &=&\!\! - (2 \gamma +i (\Delta_j-\Delta_k))  P_j P^*_k 
\\ \nonumber 
&&\!\!- \gamma C\! \sum_i\! \sqrt{p_i} (\sqrt{p_j} P_i P_k^*\! +\! \sqrt{p_k} P_j P^*_i). 
\end{eqnarray}
If $M$ is the number of velocity classes, Eq.~(\ref{pjpk}) stands for $M^2$ equations in $M^2$ variables $P_i P_k^*$. We can write them in matrix form and, in principle, invert the $M^2 \times M^2$ matrix on the right-hand side and thus solve for $P_j P^*_k$ as a linear combination of $\frac{d}{dt} (P_a P^*_b + S_a S^*_b)$ for various $a$ and $b$. Inserting this into Eq.~(\ref{effdopcav}), applying the fundamental theorem of calculus, and assuming the retrieval is complete (i.e., no excitations remain in the atoms), the retrieval efficiency can be expressed as a linear combination of $S_a(0) S^*_b(0)$, and is thus independent of control and detuning.

In Appendix \ref{sec:appretr}, we present an analogous derivation, which shows that the free-space retrieval efficiency in the presence of inhomogeneous broadening is also independent of detuning and control. Numerical calculations also show that adiabatic elimination of $P_j$, as in the homogeneously broadened case discussed in papers I and II, does not change the exact value of the efficiency.

Since the retrieval efficiency is thus independent of the exact method used for retrieval, we shall here mainly consider the fast retrieval from $S(\tilde z)$. In this case Laplace transformation in time can be used to solve the problem analytically. We will focus for the rest of this section on the free-space model. We assume that the retrieval $\pi$ pulse arrives at $\tilde t = 0$ and that it perfectly transfers $P_j(\tilde z)=0$, $S_j(\tilde z)=\sqrt{p_j} S(\tilde z)$ to $P_j(\tilde z) = i \sqrt{p_j} S(\tilde z)$, $S_j(\tilde z)=0$. After the $\pi$ pulse, we Laplace-transform Eqs.~(\ref{doppler2eqe}) and (\ref{doppler2eqp}) in time $\tilde t \rightarrow v$ and obtain
\begin{eqnarray}
\partial_{\tilde z} \eop &=& i \sqrt{d} \sum_j \sqrt{p_j} P_j,
\\
v P_j - i \sqrt{p_j} S(\tilde z) &=& - (1+ i \tilde \Delta_j) P_j + i \sqrt{d} \sqrt{p_j} \eop.
\end{eqnarray}
Solving for $P_j$ from the second equation and inserting it into the first equation, we find
\begin{equation} \label{outfastdopper}
\eop(\tilde z=1,v) = - \int_0^1 d \tilde z S(\tilde z) \sqrt{d} f(v) e^{- d f(v) (1-\tilde z)},
\end{equation}
where 
\begin{equation}\label{dopplerintegral}
f(v) = \sum_j p_j \frac{1}{1+v+i \tilde \Delta_j} = \int_{-\infty}^\infty d \tilde \Delta p(\tilde \Delta) \frac{1}{1+v+i \tilde \Delta},
\end{equation}
and where $p(\tilde \Delta)$ is the Doppler profile.

In Doppler-broadened media, the resonant optical depth is reduced by a factor of $\sim \gamma/\Delta_\textrm{I}$, where $\Delta_\textrm{I}$ is the width of the (inhomogeneous) Doppler profile. The naive expectation would therefore be that we could simply treat Doppler-broadened atoms as Doppler-free but with a reduced optical depth. As we will show below, this prescription would be correct if the broadened line shape were a Lorentzian, as considered,  e.g., in Ref.~\cite{molmerscullywelch}. For a Gaussian profile this prescription is, however, not applicable, and qualitatively different behavior is obtained.

To proceed, we first evaluate $f(v)$ for three different inhomogeneous profiles $p(\tilde \Delta)$. For a homogeneously broadened ensemble, the line shape function $p(\tilde{\Delta})$ is just a $\delta$ function:
\begin{eqnarray} \label{homo}
p(\tilde \Delta) = \delta(\tilde \Delta),
\\ \label{homo2}
f(v) = \frac{1}{1+v}.
\end{eqnarray}
If we have a Lorentzian inhomogeneous profile with $\Delta_\textrm{I}$ half width at half maximum (HWHM), we get (with $\tilde \Delta_\textrm{I} = \Delta_\textrm{I}/\gamma$)
\begin{eqnarray} \label{Loren}
p(\tilde \Delta) = \frac{\tilde \Delta_\textrm{I}}{\pi} \frac{1}{\tilde \Delta^2 + \tilde \Delta_\textrm{I}^2},
\\ \label{Loren2}
f(v) = \frac{1}{1+v+\tilde \Delta_\textrm{I}}.
\end{eqnarray}
For a Gaussian inhomogeneous profile with (rescaled by $\gamma$) standard deviation $\sigma$ (and rescaled HWHM $\tilde \Delta_\textrm{I} = \sigma \sqrt{2 \ln 2}$),
\begin{eqnarray} \label{Gauss}
p(\tilde \Delta) = \frac{1}{\sqrt{2 \pi \sigma^2}} \exp\left(-\frac{\tilde \Delta^2}{2 \sigma^2}\right),
\\ \label{Gauss2}
f(v) = \sqrt{\frac{\pi}{2 \sigma^2}} e^{\frac{(1+v)^2}{2 \sigma^2}} \textrm{erfc}\left[\frac{1+v}{\sqrt{2} \sigma}\right],
\end{eqnarray}
where the last equality assumes $\textrm{Re}[v] \geq 0$, and the complementary error function is defined as $\textrm{erfc}(x) = 1-  \pi^{-1/2} 2 \int_0^x \exp(-x'^2) d x'$. Using the definition of $\textrm{erfc}$, we can analytically continue $f(v)$ into $\textrm{Re}[v] < 0$.

If we insert the Lorentzian result for $f(v)$ from Eq.~(\ref{Loren2}) into Eq.~(\ref{outfastdopper}) and rescale $v$, we find that, compared to homogeneous broadening (Eq.~(\ref{homo2})), the Lorentzian broadening effectively just replaces $\gamma$ with $\gamma + \Delta_\textrm{I}$, which is equivalent to reducing $d$ by a factor of $1/(1+\Delta_\textrm{I}/\gamma)$ (which for $\Delta_\textrm{I} \gg \gamma$ is equal to $\gamma/\Delta_\textrm{I}$). Therefore, since we have shown that the retrieval efficiency is independent of the retrieval method, the naive rescaling of $d$ to $d\gamma/\Delta_\textrm{I}$ can indeed be used to calculate the retrieval efficiency when the broadening is Lorentzian. Similarly, the same can be shown for the cavity model.

For Doppler broadening, $p(\tilde \Delta)$ is, however, Gaussian as in Eq.~(\ref{Gauss}). Using Eqs.~(\ref{etalaplace}) and (\ref{outfastdopper}), we can write the retrieval efficiency in the form
\begin{equation}\label{doppleriterate}
\eta[S(\tilde z)] = \int_0^1 d \tilde z \int_0^1 d \tilde z' k(\tilde z,\tilde z') S(\tilde z) S^*(\tilde z')
\end{equation}
for some complicated kernel $k$. Applying the iterative technique used in Sec.~III of paper II or by directly diagonalizing $k$ on a grid, we can compute the optimal retrieval modes (i.e., the eigenvectors with the largest eigenvalues) for each $d$ and $\sigma$.

Before plotting and analyzing the optimal spin waves and the maximum efficiency obtained using Eq.~(\ref{doppleriterate}), let us discuss what we expect. Assuming $\Delta_\textrm{I} \gg \gamma$, the resonant optical depth is reduced  to $d' = d (\gamma/\Delta_\textrm{I}) \sqrt{\pi \ln 2}$ in the presence of Gaussian broadening. In contrast to retrieval with a Lorentzian profile, however, retrieval with a Gaussian profile is not equivalent to Doppler-free retrieval with reduced optical depth (as we can see by comparing Eqs.~(\ref{Gauss2}) and (\ref{homo2})). Moreover, we will show now that, with true (Gaussian) Doppler broadening, at high enough optical depth all atoms contribute coherently as if the medium were homogeneously broadened, which is the main result of this section. Although this result holds for any control, it is most easily explained in the case of fast retrieval: after the $\pi$ pulse, the spontaneous emission (or more precisely the polarization decay at a rate $\gamma$) and the dephasing due to the inhomogeneous broadening will cause the polarization $P(t) = \sum_j \sqrt{p_j} P_j(t)$ (with an initial velocity-symmetric polarization $P_j(0) = \sqrt{p_j} P(0)$) to decay as (using the original units)
\begin{eqnarray} \label{Pdecay}
P &\sim & e^{-\gamma t} \int d \Delta e^{-i \Delta t} p(\Delta/\gamma)/\gamma 
\nonumber \\
&=& \exp\left[-\gamma t - \Delta_\textrm{I}^2 t^2/(4 \ln 2)\right],
\end{eqnarray}
where we have used $p(\Delta/\gamma)$ from Eq.~(\ref{Gauss}).
Thus, losses induced by Gaussian broadening are non-Markovian. Since the time required for fast retrieval varies as $t \sim 1/(\gamma d)$ (see, for example, paper II), Doppler-induced losses become negligible compared to spontaneous emission losses for sufficiently large $d$ ($d \gg (\Delta_\textrm{I}/\gamma)^2$ or, equivalently, $d' \gg \Delta_\textrm{I}/\gamma$), and the system will behave as if there were no inhomogeneous broadening. The essential step in this last derivation is that the second moment $\langle \Delta^2\rangle$ with respect to $p(\Delta/\gamma)$ is finite, and the result is thus applicable to any inhomogeneous profile falling off faster than a Lorentzian. In contrast, for a Lorentzian profile, the optical polarization decay would be
\begin{equation}
P\sim\exp(-\gamma t-\Delta_\textrm{I} t),
\end{equation}
and we recover the effective rescaling of $\gamma $ up to $\gamma + \Delta_\textrm{I}$, which is equivalent to a simple rescaling of $d$ down to $d'$.

\begin{figure}[tb]
\begin{center}
\includegraphics[scale = 0.9]{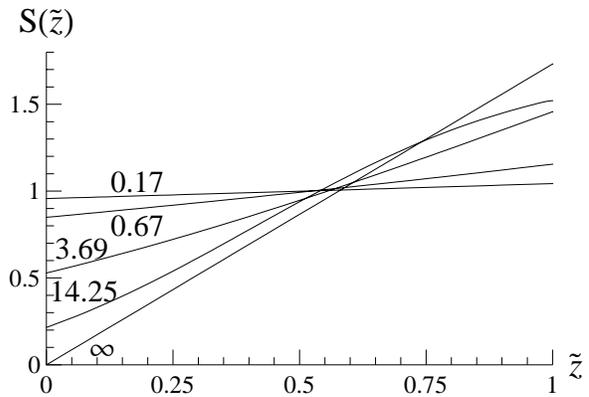}
\end{center}
\caption{Optimal spin-wave modes to retrieve from (in the forward direction) at the indicated values of $d'$ in the presence of (Gaussian) Doppler broadening with HWHM $\Delta_\textrm{I} = 88 \gamma$. \label{fig:Dopplermodes}}
\end{figure}

We now turn to the discussion of the optimal velocity-symmetric spin-wave modes $S(\tilde z)$ obtained under Gaussian broadening using Eq.~(\ref{doppleriterate}). These optimal modes are plotted (for forward retrieval) in Fig.~\ref{fig:Dopplermodes} for $d' = 0.17, 0.67, 3.69, 14.25$ and $\Delta_\textrm{I} = 88 \gamma$, which corresponds to the Rb D1 line at room temperature (assuming $2 \gamma$ is the natural linewidth). Indeed, at sufficiently high optical depth, the optimal mode approaches $S(\tilde z)= \sqrt{3} \tilde z$ (which we have derived in the homogeneously broadened case considered in paper II), since according to the argument above, Doppler broadening plays no role at sufficiently high $d$.

\begin{figure}[ht]
\begin{center}
\includegraphics[scale = 1]{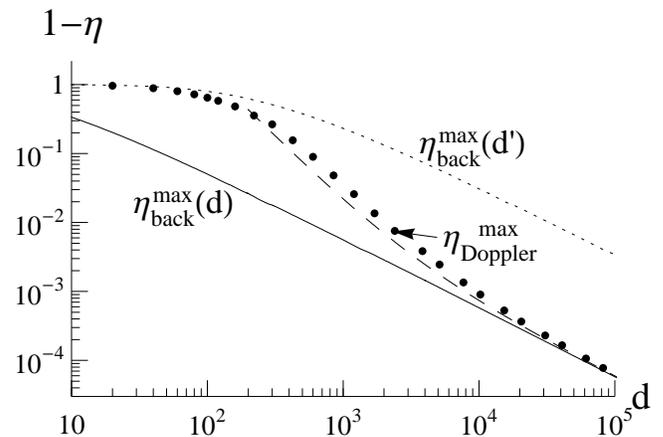}
\end{center}
\caption{Error $1-\eta$ as a function of unbroadened optical depth $d$ for different efficiencies $\eta$. $\eta^\textrm{max}_\textrm{back}$(d) (solid) is the Doppler-free maximum total efficiency for any storage followed by backward retrieval. The maximum total efficiency for storage followed by backward retrieval with Doppler broadening of HWHM $\Delta_\textrm{I} = 88 \gamma$ (circles) does not follow $\eta^\textrm{max}_\textrm{back}(d')$ (dotted), which is the efficiency one would expect from naive rescaling of resonant optical depth, but follows more closely the heuristic model $5.8 \left(\pi/(4 d'^2) + 1/d\right)$ (dashed) described in the text. \label{fig:Doppler1}}
\end{figure}

The above calculation (via Eq.~(\ref{doppleriterate})) of the optimal spin wave yields the optimal retrieval efficiency. Using the general time-reversal arguments presented in paper II, which still apply with Doppler broadening, we can, however, also use these optimal modes, to calculate the optimal efficiency for the combined process of storage followed by retrieval. The optimal symmetric modes ($S_j(\tilde z)=S(\tilde z)$ for all $j$) for retrieval, which we have found above, are all real.  Time reversal thus shows that the optimal storage into the symmetric mode, is obtained by time-reversing retrieval, and has the same efficiency as the retrieval. Note, however,  that, in general, asymmetric modes may also be excited during storage. In this section, however, we assume that the atoms rethermalize during the storage time. This washes out any amplitude on asymmetric   modes and the only relevant efficiency is the efficiency of storage onto the symmetric mode. The total maximum efficiency for storage followed by retrieval will thus be the square of the maximum retrieval efficiency (obtained as the largest eigenvalue of the kernel in Eq.~(\ref{doppleriterate})). With circles in Fig.~\ref{fig:Doppler1}, we show the maximum total efficiency for storage followed by backward retrieval for $\Delta_\textrm{I} = 88 \gamma$. The solid line in Fig.~\ref{fig:Doppler1} is the Doppler-free maximum efficiency $\eta^\textrm{max}_\textrm{back}(d)$ from Fig.~4 in paper II. The dotted line is the efficiency one would naively expect from a simple rescaling of the resonant optical depth, $\eta^\textrm{max}_\textrm{back}(d')$, where $d' = d (\gamma/\Delta_\textrm{I}) \sqrt{\pi \ln 2}$. The dashed line is $5.8 \left(\pi/(4 d'^2) + 1/d \right)$, which approximates the error fall-off of the points obtained from the full numerical optimization of Eq.~(\ref{doppleriterate}) (circles) reasonably well and comes from the following heuristic model: $t = 1/(\gamma d)$ is inserted into Eq.~(\ref{Pdecay}), the exponential is expanded to first order, and a prefactor of $5.8$ is introduced to match the Doppler-free error at large $d$, which is $\sim 5.8 /d$, as found in paper II.

Above, we found the optimal spin-wave modes for retrieval and the optimal retrieval efficiency by computing the eigenvector with the largest eigenvalue of the kernel in Eq.~(\ref{doppleriterate}). As in the homogeneously broadened case discussed in paper II, one can, in fact, interpret the iterations used in finding the largest eigenvalue as the iterations of retrieval followed by storage of time-reversed output and control pulses. To show this, we note that the retrieval in Eq.~(\ref{outfastdopper}) inverse-Laplace-transformed back to $\tilde t$ and the storage equation, which is obtained using the same steps, can be written as
\begin{eqnarray} \label{DopES} 
\eop_\textrm{out}(\tilde t) &=& \int_0^1 d \tilde z \; \tilde m\!\!\left(\tilde t, \tilde z\right) S(1-\tilde z),
\\ \label{DopSE}
S(\tilde z, \tilde T) &=& \int_0^{\tilde T} d \tilde t \; \tilde m\!\!\left(\tilde T - \tilde t, \tilde z\right) \eop_\textrm{in}(\tilde t), 
\end{eqnarray}
for some function $\tilde m$. Since these equations satisfy the general time-reversal form of Eqs.~(35) and (36) in paper II, the same results as in the homogeneously broadened case apply. In particular, using Eqs.~(\ref{DopES}) and (\ref{DopSE}), one can check that the maximization of retrieval efficiency through the iterative integration of the kernel in Eq.~(\ref{doppleriterate}) is equivalent to retrieval followed by time reversal and storage. From Eqs.~(\ref{DopES}) and (\ref{DopSE}), it also follows that, in order to optimize fast storage followed by fast forward retrieval, one should start with any trial input mode, store it, retrieve forward, time-reverse the whole process, and then iterate till convergence is reached, exactly as in the homogeneously broadened case of paper II. 

It is worth noting that a connection between time reversal and optimal photon storage in the photon-echo technique was first made for the case of ideal, reversible storage in Refs.~\cite{moiseev01, kraus06, photonecho}. In the present paper and in Refs.~\cite{gorshkov07, paperI, paperII}, we extend this connection to a wide range of storage techniques in $\Lambda$-type media and show that optimal storage is intimately connected with time reversal, even when the dynamics of the system are not completely reversible, and when the ideal unit efficiency cannot be achieved.

We have thus shown that time-reversal iterations can be used to optimize storage followed by retrieval not only in the case of homogeneous broadened media discussed in paper I, paper II, and Ref.~\cite{gorshkov07}, but also in the case of Doppler-broadened media. As explained in paper II, such time-reversal iterations 
not only constitute a convenient mathematical tool; they can, in fact, be used experimentally to find the optimal input modes. In particular, as explained in paper II, since the envelope $\eop$ of the quantum light mode obeys the same equations of motion as classical light pulses, one can first use the iterative procedure to optimize the storage of classical light pulses, which can be easily measured and reversed, and then directly apply this knowledge to the storage of quantum states of light. In fact, the first experiment on time-reversal-based optimization has already been successfully carried out for classical light \cite{novikova07} and confirmed the validity of the suggested procedure. 

\section{Inhomogeneous Broadening without Redistribution between Frequency Classes during the Storage Time \label{sec:inhom}}

In the previous section, we treated the case when inhomogeneously broadened atoms redistribute themselves among different frequency classes during the storage time, which is the case in Doppler-broadened atomic vapors for sufficiently long storage times. This redistribution, however, does not take place in some other possible experimental realizations, e.g., in Doppler-broadened atomic vapors with short storage times or in solid state media. Therefore, in this section, we consider what happens when the redistribution among frequency classes does not take place. 

In the case of fast storage and retrieval, provided the $\pi$ pulse is applied at a sufficiently high power, it does not matter whether the two optical transitions are broadened independently or not, i.e., whether the $|s\rangle-|g\rangle$ transition is broadened. However, in the case of adiabatic storage and retrieval the assumption that the control and the quantum field are always in two-photon resonance is crucial. Although the only regime of storage and retrieval in inhomogeneously broadened media we will consider in this section is the fast regime, the analysis will also be extendable to the adiabatic limit provided the $|s\rangle-|g\rangle$ transition is homogeneously broadened. In this case, Eqs.~(\ref{doppler2eqe})-(\ref{doppler2eqs}) apply without modification. The proof in Appendix \ref{sec:appretr} that the retrieval efficiency is independent of the detuning and the control, therefore, also applies. 

Using the solution technique based on the Laplace transformation in time introduced in Sec.~\ref{sec:Doppler}, one can show that when storage is followed by forward retrieval and the inhomogeneous profile is Lorentzian, it actually does not matter whether the atoms redistribute themselves among different frequency classes during the storage time or keep their frequencies unchanged: the same efficiency and output field are obtained. In this case, the results from paper II about homogeneous broadening are directly applicable if one replaces $d$ by $d'$. This is, however, not true for backward retrieval with a Lorentzian inhomogeneous profile or for retrieval in either direction with a Gaussian inhomogeneous profile. To obtain the efficiency in this situation, it is therefore necessary to take into account the fact that the transition frequency of each individual atom is the same during both storage and retrieval. Furthermore, by controlling and reversing the inhomogeneous broadening, one can even achieve rephasing of atomic excitations and, in fact, attain an increase in total efficiency relative to an unbroadened case \cite{kroll05,kraus06}. An exhaustive study of the problem of storage followed by retrieval in media with no redistribution between frequency classes during the storage time is beyond the scope of this paper. Here we restrict ourselves only to the investigation of fast storage followed by fast backward or forward retrieval in such media. We also include the possibility of reversing the inhomogeneous profile during the storage time as suggested in Refs.~\cite{kroll05,kraus06}. In particular, in Sec.~\ref{sec:InhomSol}, we set up the equations for the problem of fast storage followed by fast retrieval in \textit{either} direction with and without the reversal of the inhomogeneous profile. In Secs.~\ref{sec:InhomBackOpt} and \ref{sec:CRIB}, we then discuss the results that these equations yield for the cases of storage followed by \textit{backward} retrieval without and with the reversal of the inhomogeneous profile, respectively. 

\subsection{Setup and solution \label{sec:InhomSol}}

In this section, assuming that the redistribution between frequency classes takes place neither during the processes of storage and retrieval nor during the storage time, we set up and solve the problem of fast storage followed by fast retrieval in the forward or backward direction with or without the reversal of the inhomogeneous broadening. Any storage with no redistribution between frequency classes (not only the fast limit) can be computed numerically with discrete frequency classes for any kind of inhomogeneous profile and any control. To do this, one can just use Eqs.~(\ref{doppler2eqe})-(\ref{doppler2eqs}) for both storage and retrieval and, depending on the direction of retrieval and on whether the inhomogeneous profile is reversed during the storage time, make the appropriate modification to the stored spin waves $S_j(\tilde z, \tilde T)$ before retrieving it. However, as we now show, the case of fast storage and fast retrieval can also be solved almost completely analytically using Laplace transformation in time, $\tilde t \rightarrow v$ \cite{fastnote}. Before the storage $\pi$ pulse is applied, $\tilde \Omega = 0$ in Eq.~(\ref{doppler2eqp}). Then for the fast storage of a resonant input mode $\eop_\textrm{in}(\tilde t)$, Eqs.~(\ref{doppler2eqe}) and (\ref{doppler2eqp}) can be solved to give
\begin{equation} \label{fastP}
P_j(\tilde z,v) = i \sqrt{d} \frac{\sqrt{p_j}}{1+v+i \tilde \Delta_j} \eop_\textrm{in}(v) e^{-d \tilde z f(v)},
\end{equation}
where $v$ is the Laplace variable, and $f(v)$ is defined in Eq.~(\ref{dopplerintegral}) and is computed for various inhomogeneous profiles in Eqs.~(\ref{homo})-(\ref{Gauss2}).

To find the initial conditions for the subsequent retrieval, we take the inverse Laplace transform $u \rightarrow \tilde t = \tilde T$ of Eq.~(\ref{fastP}) and multiply $P_j$ by $-1$ to account for the two $\pi$ pulses (i.e., the storage and retrieval $\pi$ pulses). If we are interested in backward retrieval, $P_j(\tilde z)$ should be flipped to $P_j(1-\tilde z)$. If we are interested in reversing the inhomogeneous profile, the frequency classes should be reversed. Thus, for example, for backward retrieval with the reversal of inhomogeneous broadening, the initial condition for retrieval is $P_j(\tilde z,\tilde T_\textrm{r})=-P_{-j}(1-\tilde z,\tilde T)$. Using Eqs.~(\ref{doppler2eqe}) and (\ref{doppler2eqp}) with $\tilde \Omega = 0$, we can then implement fast retrieval. The time Laplace transform of the output field can then be found to be equal to  
\begin{equation} \label{eoutv}
\eop_{out}(v) \!=\! \mathcal{L}^{-1}\!\left\{A(v,v') B(v,v') \eop_\textrm{in}(v')\right\}_{v' \rightarrow \tilde T},
\end{equation}
where $\mathcal{L}^{-1}$ indicates that we should take the inverse Laplace transform $v' \rightarrow \tilde T$, $A(v,v')$ depends on the direction of retrieval and is given by
\begin{eqnarray}
A(v,v')&=& \frac{e^{-d (f(v) + f(v'))} - 1}{f(v)+f(v')},
\\
A(v,v')&=& \frac{e^{-d f(v)} - e^{-d f(v')}}{f(v)-f(v')}
\end{eqnarray}
for backward and forward retrieval, respectively, and $B(v,v')$ depends on whether the inhomogeneous profile is reversed or not and is given by
\begin{eqnarray}
B(v,v')&=& \frac{f(v)+f(v')}{2+v+v'},
\\
B(v,v')&=& \frac{f(v)-f(v')}{v'-v}
\end{eqnarray}
for reversed and not reversed cases, respectively.

For homogeneous broadening (Eqs.~(\ref{homo}) and (\ref{homo2})) and for a Lorentzian inhomogeneous profile (Eqs.~(\ref{Loren}) and (\ref{Loren2})), the inverse Laplace transforms $v' \rightarrow \tilde T$ and $v \rightarrow \tilde t$ can be taken analytically in terms of Bessel functions and convolutions. The case of homogeneous broadening has been studied in paper II, while the analytical answer for the Lorentzian case is too complicated to yield any significant insight. For a Gaussian inhomogeneous profile (Eqs.~(\ref{Gauss}) and (\ref{Gauss2})), the inverse Laplace transforms have to be taken numerically. In all three cases, the efficiency can be computed via Eq.~(\ref{etalaplace}) without taking the $v \rightarrow \tilde t$ inverse Laplace transform.

If we are interested not only in computing the total efficiency of storage followed by retrieval for some given input photon mode, but also in maximizing the efficiency with respect to the input mode shape, we can again take advantage of time reversal. We will show now that in all four cases (i.e., either of the two retrieval directions, with or without the reversal of broadening), the optimal input pulse shape can be found by starting with any trial input, carrying out storage followed by retrieval, then time-reversing the output, and iterating the procedure till convergence is reached. To begin the proof, we note that $A(v,v')$ and $B(v,v')$ are symmetric with respect to the exchange of the two arguments. Therefore, Eq.~(\ref{eoutv}) and the convolution theorem for Laplace transforms imply that we can write
\begin{equation}
\eop_\textrm{out}(\tilde t) = \int_0^{\tilde T} d \tilde t' \eop_\textrm{in}(\tilde t') m'(\tilde t, \tilde T - \tilde t')
\end{equation} 
for some function $m'$ that is symmetric with respect to the exchange of its two arguments. One can also check that $m'$, and hence the optimal input mode, are real. Assuming, therefore, a real $\eop_\textrm{in}(\tilde t)$, the total efficiency is
\begin{equation}
\eta = \int_0^{\tilde T} d \tilde t \int_0^{\tilde T} d \tilde t' \eop_\textrm{in}(\tilde t) \eop_\textrm{in}(\tilde t') k_\textrm{tot}(\tilde t, \tilde t'), 
\end{equation}
 where the kernel $k_\textrm{tot}(\tilde t, \tilde t')$ (the subscript ``tot" stands for the total efficiency, i.e., storage followed by retrieval) is given by
\begin{equation}
k_\textrm{tot}(\tilde t, \tilde t')= \int_0^{\tilde T} d \tilde t'' m'(\tilde t'', \tilde T - \tilde t) m'(\tilde t'', \tilde T - \tilde t'),
\end{equation} 
where we assumed $\tilde T$ is sufficiently large that the interval $[0,\tilde T]$ includes the whole retrieved pulse. To find the optimal $\eop_\textrm{in}(\tilde t)$, one can thus start with a trial input mode $\eop_1(\tilde t)$ and iterate the action of the kernel according to
\begin{equation}
\eop_2(\tilde t') = \int_0^{\tilde T} d \tilde t  \eop_1(\tilde t) k_\textrm{tot}(\tilde t, \tilde t').
\end{equation}
Using the symmetry of $m'$, one can immediately see that this iteration is equivalent to carrying out storage followed by retrieval, time-reversing the output, repeating the procedure, and time-reversing the output again. We will use these time-reversal iterations in the following sections to compute the optimal input modes. It is important to note that, as we have explained at the end of Sec.~\ref{sec:dopretst}, the time-reversal iterations that we have just described not only constitute a convenient mathematical tool; they can, in fact, be used experimentally to find the optimal input modes.

In this section, we set up and solved the problem of fast storage followed by fast backward or forward retrieval with or without the reversal of the inhomogeneous profile. In the next two sections, we would like to analyze these solutions for the case of backward retrieval.

\subsection{Storage followed by backward retrieval \label{sec:InhomBackOpt}}

In this section, using the results of Sec.~\ref{sec:InhomSol}, we would like to study the efficiency of fast storage followed by fast backward retrieval in inhomogeneously broadened media without redistribution between velocity classes, and without the reversal of the inhomogeneous broadening during the storage time. This problem is motivated by solid state implementations such as the one described in Refs.~\cite{kroll04, afzelius06}, where the line shape is created by pumping back some atoms from a broad spectral hole feature. In this situation, one can consider what happens when one expands the spectral region and pumps more and more atoms into the absorptive feature. In this case, the resonant optical depth $d'$ is expected to be independent of the width of the spectral feature that is pumped back into the absorption profile, whereas the unbroadened optical depth $d$ would increase with an increasing number of atoms being pumped back. From the analysis of Sec.~\ref{sec:Doppler}, we expect the behavior to be different depending on whether the line shape falls off as a Lorentzian or faster. In particular, we found in Sec.~\ref{sec:Doppler} that for a well-localized inhomogeneous line (e.g., a Gaussian), the losses due to dephasing are non-Markovian and, therefore, scale as $1/d'^2$, in contrast to the losses due to the exponential polarization decay with the rate $\gamma$, which scale as $1/d$. As another application of this idea, we show in this section that, for a Gaussian inhomogeneous profile, by increasing $d$, the error during fast storage followed by fast backward retrieval can be lowered from the $1/d'$ scaling for the homogeneously broadened line ($d = d'$) to the $1/d'^2$ scaling for sufficiently large $d$. Experimentally, $d'$ can be quite large ($d' \approx 50$ should be feasible by using a sufficiently high impurity concentration and a sufficiently long sample length \cite{kroll05,afzelius06}), so that this could potentially yield a very high efficiency.

We begin the analysis by using the time reversal iterations suggested at the end of Sec.~\ref{sec:InhomSol} to compute the optimal input pulse and the maximum possible efficiency at various values of $d$ and $d'$ assuming the inhomogeneous profile is Gaussian (Eq.~(\ref{Gauss})). To compute the standard deviation $\sigma$ of the Gaussian profile from the values of $d$ and $d'$, we use the following relationship:
\begin{equation}
d' = d \times \int_{-\infty}^{\infty} d \tilde \Delta p(\tilde \Delta) \frac{1}{1 + \tilde \Delta^2} = \sqrt{\frac{\pi}{2}} \frac{d}{\sigma} e^{\frac{1}{2 \sigma^2}} \textrm{erfc}\left[\frac{1}{\sqrt{2} \sigma} \right],
\end{equation} 
where the second factor in the integrand is the homogeneous line shape of HWHM equal to 1 (recall that our frequencies are rescaled by $\gamma$).

\begin{figure}[ht]
\begin{center}
\includegraphics[scale = 0.9]{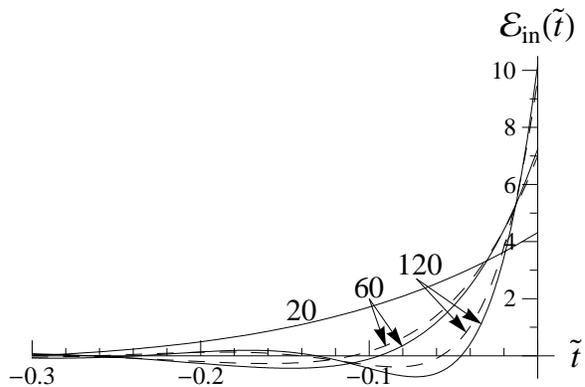}
\end{center}
\caption{The solid lines show the optimal input modes for $d'=20$ and $d=20, 60$, and $120$ (the value of $d$ is indicated on the plot) for storage followed by backward retrieval in a medium with a Gaussian inhomogeneous profile, without the reversal of the inhomogeneous broadening, and without averaging over frequency classes during the storage time. The dashed lines show the corresponding optimal modes for $d'=d$, i.e., for the case of no inhomogeneous broadening. \label{fig:OptimalInhomModes}}
\end{figure}

In Fig.~\ref{fig:OptimalInhomModes}, we show (solid lines) the optimal input modes for $d' = 20$ and $d = 20, 60$, and $120$. We see that these optimal input modes (for this value of $d'$) are very similar to the optimal input modes (dashed lines) for the same $d$ but without the inhomogeneous broadening (i.e., $d' = d$). The optimal modes thus have a duration of approximately $1/(d \gamma)$, as discussed in Sec.~VII of paper II, and consist of a roughly triangular pulse preceded by a few ``wiggles." These wiggles can be traced back to the zeroth-order Bessel function of the first kind (i.e., $J_0$) in Eq.~(48) in paper II and can be associated with Rabi oscillations between the electric field $\eop$ and the optical polarization $P$. In Fig.~\ref{fig:OptimalInhomModes}, we assume that the storage $\pi$ pulse is applied at $\tilde t = 0$. Although the true optimal input pulses extend to $\tilde t = -\infty$, they can, in practice, be truncated after about two or three wiggles without decreasing the efficiency by more than $10^{-4}$.

\begin{figure}[bt]
\begin{center}
\includegraphics[scale = 0.9]{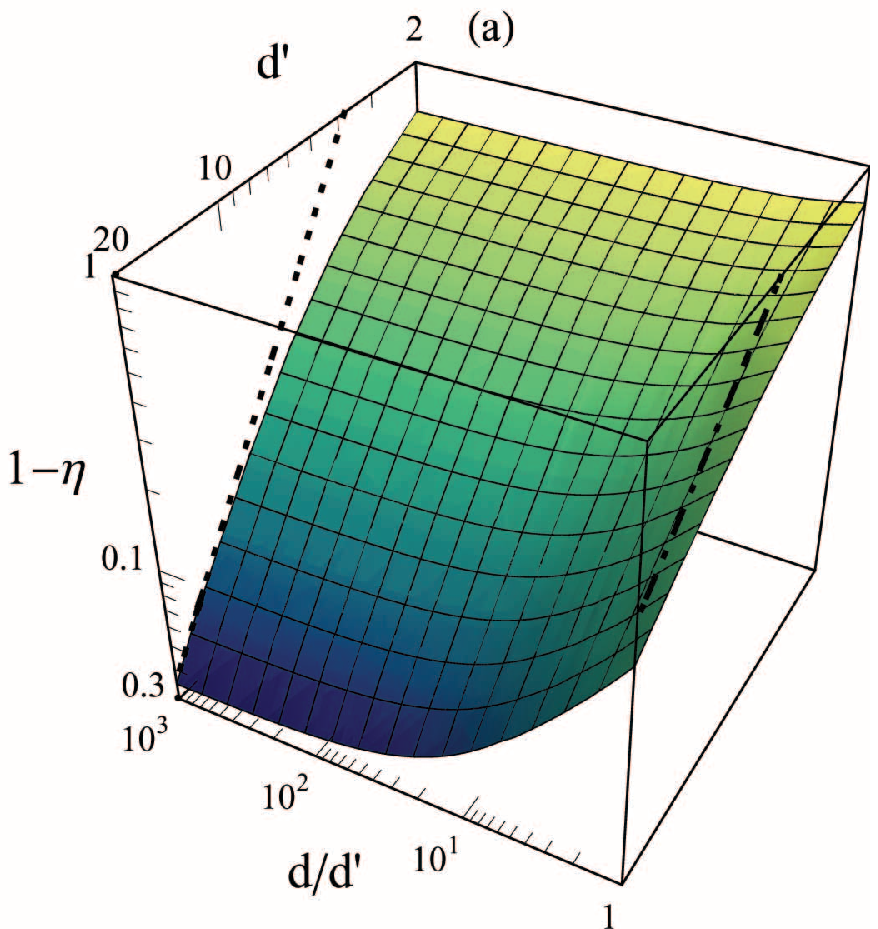}
\includegraphics[scale = 1]{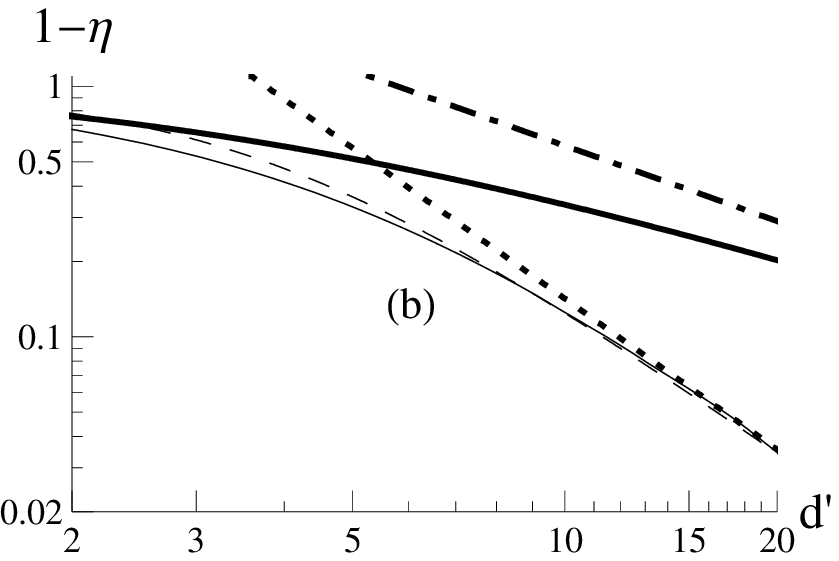}
\end{center}
\caption{(Color online) (a) As a function of $d'$ and $d/d'$, the optimal (smallest) error of fast storage followed by fast backward retrieval (surface). $5.8/d'$ (dash-dotted line) is the limiting behavior, found in paper II, of the homogeneous error ($d/d' = 1$) for large enough $d$ ($= d'$). $14.2/d'^2$ (dotted) approximates the $d'$-limited error ($d/d' \rightarrow \infty$) near $d' \sim 10-20$ and, probably, above. (b) As a function of $d'$, dash-dotted and dotted lines are the same as in (a). The thick and thin solid lines are the $d/d' = 1$ and $d/d' \rightarrow \infty$ error curves, respectively, which can be read out from the surface plot in (a).  $1/(1+14.2/d'^2)$ (dashed line) is a heuristic curve that matches the $d'$-limited error (thin solid line) very well. \label{fig:OptimalBackInhom}}
\end{figure}

To verify the prediction that, for sufficiently large $d$, the error should be limited by $1/d'^2$, we computed the optimal (smallest) error by optimizing with respect to the input mode at different values of $d'$ and $d$. The optimization was done numerically using the time-reversal iterations suggested at the end of the previous section. Figure \ref{fig:OptimalBackInhom}(a) shows a log-log surface plot of this optimal (smallest) error $1-\eta$ as a function of $d'$ and $d/d'$. As expected, we find that, for any fixed $d'$, the error is very well approximated by $c_1 + c_2/d$ at large $d$, where the constants $c_1(d')$ and $c_2(d')$ depend on $d'$. $c_2(d')$ is of order unity and increases approximately linearly from about $0.2$ at $d' = 2$ to $4.2$ at $d' = 20$. $c_1(d')$ represents the $d'$-limited error, i.e., the limit $d/d' \rightarrow \infty$, when the $1/d$ error becomes negligible. This $d'$-limited error can be seen at the $d/d' = 10^3$ edge of the box in Fig.~\ref{fig:OptimalBackInhom}(a) and is also plotted separately as a thin solid line in the log-log plot in Fig.~\ref{fig:OptimalBackInhom}(b). The dotted line in Figs.~\ref{fig:OptimalBackInhom}(a) and \ref{fig:OptimalBackInhom}(b) is $14.2/d'^2$ and is shown to indicate that the $d'$-limited error indeed scales as $1/d'^2$. To include the requirement that the efficiency drops to zero at $d' = 0$ and to reproduce the $14.2/d'^2$ dependence that we see at larger $d'$, a dashed heuristic curve $1/(1+14.2/d'^2)$ is shown in Fig.~\ref{fig:OptimalBackInhom}(b). We see that it matches the exact value of the $d'$-limited error (thin solid line) very well. 

We compare this $d'$-limited error to the smallest possible error for the homogeneously broadened case $d = d'$, which can be seen at the $d/d' = 1$ edge of the box in Fig.~\ref{fig:OptimalBackInhom}(a) and which is also plotted separately as a thick solid line in Fig.~\ref{fig:OptimalBackInhom}(b). This thick solid line is the same as the solid curve in Fig.~4 of paper II and in Fig.~\ref{fig:Doppler1} of the present paper. Since we know from the discussion in paper II that at high enough values of $d$ ($= d'$), this error scales as $\sim 5.8/d$, we plot this $5.8/d$ scaling as the dash-dotted line in Figs.~\ref{fig:OptimalBackInhom}(a) and \ref{fig:OptimalBackInhom}(b). We thus see that, for a fixed $d'$, by increasing  the number of atoms in the absorption line so that we go from $d/d' = 1$ to $d/d' \rightarrow \infty$ (in such a way that the inhomogeneous line is Gaussian), one can significantly lower the optimal (smallest) error from a $1/d'$ homogeneous error to a $1/d'^2$ inhomogeneous-broadening-limited error. Although we have investigated only backward retrieval, we expect that for the case of optimal storage followed by forward retrieval, the same result will apply, and that the optimal $1/d'$ homogeneous error can also be reduced to a $d'$-limited $1/d'^2$ error. We also expect these error scalings to hold not only for the Gaussian inhomogeneous profile but also for any inhomogeneous profile (such as, for example, a square profile) whose tails fall off faster than Lorentzian. 

To summarize, we have shown that, even without any additional experimental requirements, the storage and retrieval by two fast $\pi$ pulses may be a very promising route to a quantum memory. Indeed, if the absorptive feature is sufficiently well localized (falling off faster than a Lorentzian), the error may be limited to $\sim 1/d'^2$. In a practical realization, it would, however, be necessary to include other imperfections such as imperfect $\pi$ pulses, imperfect synchronization between the storage $\pi$ pulse and the input pulse, and the limitations associated with the creation of an absorption line. Furthermore, the optimal scenario described in this section applies only to a single input pulse shape at any given values of $d$ and $d'$. For other pulse shapes, it may be advantageous to combine the ideas of this section with the reversible broadening of the next section and  optimize with respect to both the width (and shape) of the original nonreversible line and the width (and shape) of the extra reversible broadening.  

\subsection{Storage followed by backward retrieval with the reversal of the inhomogeneous profile \label{sec:CRIB}}

In the previous section, we assumed that the inhomogeneous broadening had a fixed distribution, e.g., due to different environments for each individual atom. In this section, using the results of Sec.~\ref{sec:InhomSol}, we investigate the possibility of improving the efficiency of fast storage followed by fast backward retrieval by adding and reversing inhomogeneous broadening in an originally homogeneously broadened medium. 

The possible advantage of inhomogeneous broadening was first considered in Ref.~\cite{moiseev01}, where it was noted that Doppler broadening automatically reverses during backward retrieval. This results in a reversal of the dephasing occurring during storage and gives rise to photon echo. In Refs.~\cite{kroll05,kraus06} it was then suggested that similar effects could be realized in solid state systems. Under the name of controlled reversible inhomogeneous broadening (CRIB), the authors of Ref.~\cite{kraus06} suggest implementing the equivalent of the fast storage protocol considered in paper II, but in addition they suggest controllably adding inhomogeneous broadening to the transition and then reverse the broadening during retrieval to obtain a rephasing  \cite{nopipulse}. Several experimental groups are currently working on the realization of CRIB \cite{kroll05,manson06,afzelius06}. We will show that, although the introduction of reversible inhomogeneous broadening can improve the efficiency of fast storage of a single pulse, the improvement relative to the fast storage technique without inhomogeneous broadening is limited. We will also show that CRIB can perform slightly better than optimal adiabatic storage in a homogeneously broadened medium (discussed in Secs.~VI B and VI C of paper II), but only for short pulses, for which the adiabaticity condition $T d \gamma \gg 1$ is not satisfied.

For concreteness, we consider the storage of a resonant Gaussian-like pulse of variable time duration $T$ defined, as in Eq.~(40) in paper II, by  
\begin{equation} \label{Gaussiancopy}
\eop_\textrm{in}(\tilde t) = A(e^{-30 (\tilde t/\tilde T - 0.5)^2} - e^{-7.5})/\sqrt{\tilde T}
\end{equation}
and shown in Fig.~3 of paper II. $A \approx 2.09$ is a normalization constant. We also restrict ourselves to the situation where the medium is initially homogeneously broadened. Although this is often a good approximation, the hole-burning technique in the solid state \cite{kroll04}, for example, will always result in some residual inhomogeneous broadening of the prepared line, something that one may have to take into account in a complete assessment of the performance of CRIB.

\begin{figure}[htb]
\begin{center}
\includegraphics[scale = 0.9]{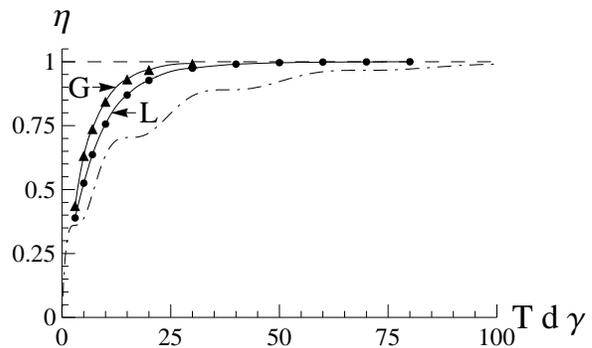}
\end{center}
\caption{Total efficiency of fast storage followed by fast backward retrieval as a function of $T d \gamma$ in the limit when $T \gamma \rightarrow 0$, $d \rightarrow \infty$, but $T d \gamma$ is finite. The curves show the efficiency  without inhomogeneous broadening (dash-dotted line) and with (solid lines) Gaussian (G) or Lorentzian (L) inhomogeneous profiles optimized with respect to the inhomogeneous width. The horizontal dashed line is 1. \label{fig:CRIB2}}
\end{figure}

To investigate the performance of CRIB, we use the techniques described in Sec.~\ref{sec:InhomSol}. Let us initially assume $T \gamma \ll 1$, so that the decay can be  ignored. As a test case, we take the resonant Gaussian-like pulse of Eq.~(\ref{Gaussiancopy}) and implement fast storage of it (with a $\pi$ pulse at $t = T$) followed by fast backward retrieval. From the discussion of CRIB \cite{kraus06}, it is expected that  adding some broadening and thereby  increasing the width of the  absorption line, at the expense of a decreased resonant optical depth, may increase the total efficiency. 
Excessive broadening will, however, make the medium transparent and decrease  the efficiency.  At each value of $T d \gamma$, there is thus an optimal inhomogeneous width. In Fig.~\ref{fig:CRIB2}, the total efficiency is plotted  as a function of $T d \gamma $, for the homogeneously broadened case (dash-dotted line) and for an inhomogeneously broadened medium (solid lines) with Gaussian (G) or Lorentzian (L) inhomogeneous profiles optimized with respect to the inhomogeneous width. The horizontal dashed line is 1. These curves are calculated by numerically computing, at each $T d \gamma$, the efficiency as a function of the inhomogeneous width and then finding the width that gives the maximum efficiency at the chosen value of $T d \gamma$. Note that, even though we neglect the decay ($\gamma T \ll 1$), the quantity $T d \gamma = g^2 N T L/c$ may still attain a non-negligible value for a sufficiently high optical depth $d\gg1$. In other words, the curves in Fig.~\ref{fig:CRIB2} represent the limit $T \gamma \rightarrow 0$ and $d \rightarrow \infty$ with finite $T d \gamma$.
Below $T d \gamma \sim 1$, adding broadening only lowers the efficiency, so that the optimal curves (solid) join the unbroadened curve (dash-dotted).
 We see, however, that at intermediate values of $T d \gamma \sim 10$, introducing reversible inhomogeneous broadening can increase the efficiency. 
 The gain in efficiency is, however, limited. In an experimental realization, one should therefore evaluate whether this gain justifies the additional experimental efforts needed to implement the controlled reversible inhomogeneous broadening.  

\begin{figure}[b]
\begin{center}
\includegraphics[scale = 0.9]{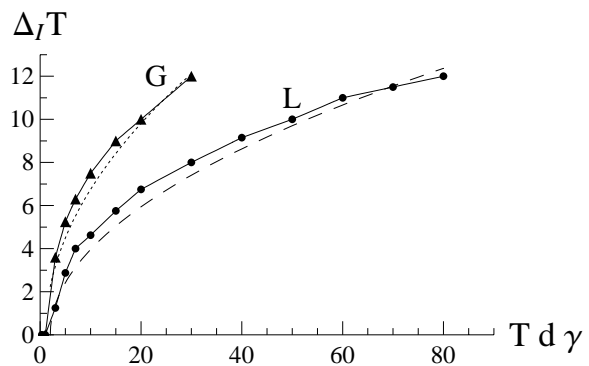}
\end{center}
\caption{$\Delta_\textrm{I} T$ as a function of $T d \gamma$, where $\Delta_\textrm{I}$ is the optimal HWHM of the Gaussian (G) or Lorentzian (L) inhomogeneous profile. The points are connected with straight lines to guide the eye. The dashed and dotted lines are $1.4(T d \gamma - 2)^{1/2}$ and $2.25(T d \gamma - 2)^{1/2}$, respectively. \label{fig:CRIB3}}
\end{figure}

In Fig.~\ref{fig:CRIB3}, we plot $\Delta_\textrm{I} T$ as a fuction of $T d \gamma$, where $\Delta_\textrm{I}$ is the optimal HWHM of the Gaussian (G) or Lorentzian (L) inhomogeneous profile used to construct  Fig.~\ref{fig:CRIB2}. The points are connected with straight lines for better visibility. The dashed and dotted lines are $1.4(T d \gamma - 2)^{1/2}$ and $2.25(T d \gamma - 2)^{1/2}$, respectively, which indicates that the optimal inhomogeneous HWHM $\Delta_\textrm{I}$ scales approximately as $\sqrt{T d \gamma}/T$, which is different from the naive guess $\Delta_\textrm{I} \sim 1/T$. We also see that, at a given $T d \gamma$, the optimal Gaussian profile is wider; and, moreover, from Fig.~\ref{fig:CRIB2}, we see that optimal Gaussian broadening gives greater efficiency than optimal Lorentzian broadening. One could imagine that these two results are the consequence of the Gaussian frequency profile of the input pulse. However, we ran an equivalent simulation with an input pulse that has a Lorentzian spectrum; and also in this case the optimal Gaussian profile is wider than the optimal Lorentzian profile and the optimal Gaussian broadening gives a greater efficiency than the optimal Lorentzian broadening. These results reflect the fact that the storage we are considering is a dynamical process and is therefore not accurately described by its continuous-wave absorption profile. We believe that the advantage of Gaussian broadening over Lorentzian comes from the fact that, as shown in Sec.~\ref{sec:Doppler}, due to their non-Markovian nature, the dephasing losses associated with Gaussian (and, hence, fast-falling) inhomogeneous broadening are smaller than the losses associated with Lorentzian broadening. 

\begin{figure}[ht]
\begin{center}
\includegraphics[scale = 0.9]{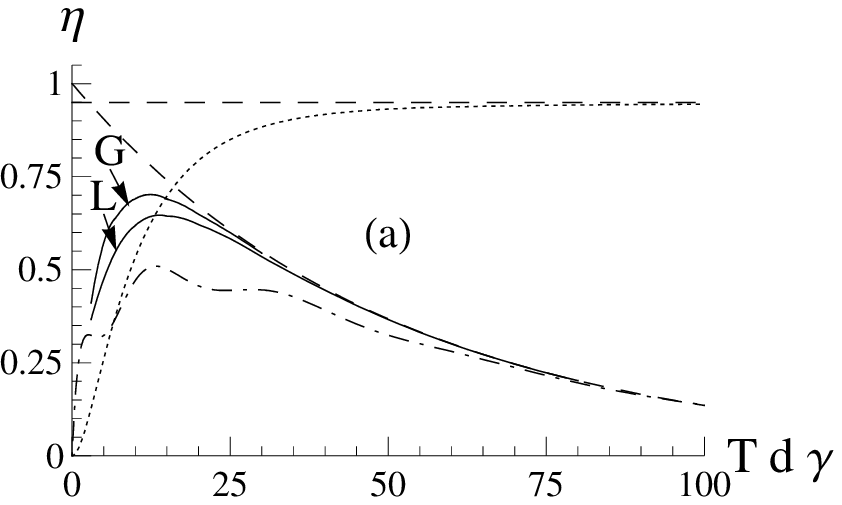}
\includegraphics[scale = 0.9]{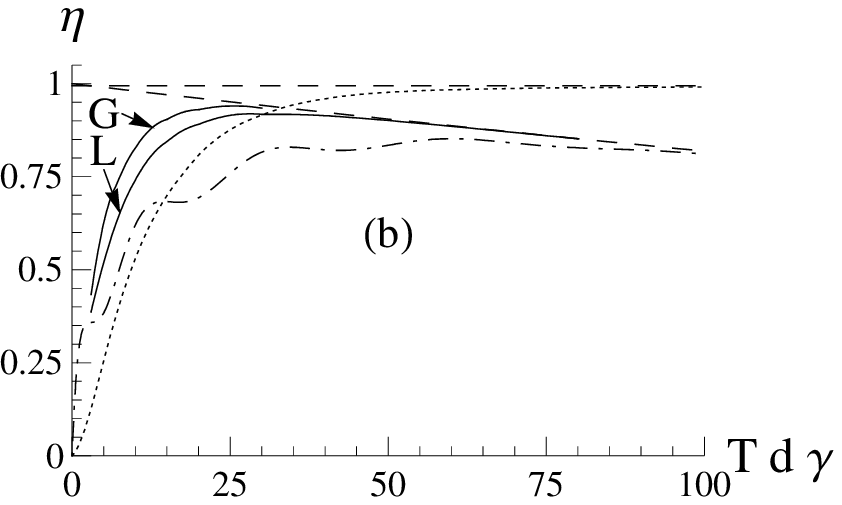}
\end{center}
\caption{Storage and backward retrieval with and without reversible inhomogeneous broadening at finite optical depth  $d = 100$ (a), and (b) $1000$. 
The curves show
the total efficiency of storage followed by backward retrieval of the resonant Gaussian-like pulse of duration $T$ of Eq.~(\ref{Gaussiancopy}) (shown in Fig.~3 of paper II) with ``optimal" resonant adiabatic storage (dotted) or with fast storage without inhomogeneous broadening (dash dotted) or with reversible optimal-width Lorentzian (L) or Gaussian (G) broadening (solid). The horizontal dashed line is the optimal adiabatic efficiency while the second dashed line is $\exp(-2 \gamma T)$, by which fast efficiencies are rescaled relative to the $T \gamma \rightarrow 0$ limit of Fig.~\ref{fig:CRIB2}. \label{fig:CRIB4}}
\end{figure}

To investigate the performance of the storage and retrieval protocol at a finite optical depth, we now relax the assumption $\gamma T \ll 1$. With the pulse considered here (Eq.~(\ref{Gaussiancopy})), the effect of the spontaneous emission $\gamma$ on fast storage with CRIB can be estimated based on simple arguments: the pulse is symmetric around $T/2$ so the excitation spends on average a time $T/2$ in the sample both during storage and retrieval. In each of these processes, the efficiency is therefore decreased by $\left[\exp(-\gamma T/2)\right]^2$ so that the total efficiency is reduced by approximately $\exp(-2 \gamma T)$. We have explicitly verified this simple estimate for a few cases and found it to be true both with and without broadening. The optimization of broadening without decay therefore gives the same optimal inhomogeneous width as with decay. Since we have in Sec.~VI D of paper II calculated adiabatic efficiencies for the same pulse shape, we can now compare the performance of fast storage with and without CRIB to adiabatic storage. In Fig.~\ref{fig:CRIB4}(a), with $d=100$, we compare the  storage of the Gaussian-like pulse of duration $T$ of Eq.~(\ref{Gaussiancopy}) (shown in Fig.~3 of paper II) followed by backward retrieval using ``optimal" adiabatic storage (dotted) or using fast storage without inhomogeneous broadening (dash dotted) or with reversible optimal-width Lorentzian (L) or Gaussian (G) broadening (solid). The horizontal dashed line is the optimal adiabatic efficiency, while the second dashed line is $\exp(-2 \gamma T)$, by which fast efficiencies are rescaled relative to the $T \gamma \rightarrow 0$ limit of Fig.~\ref{fig:CRIB2}. Figure \ref{fig:CRIB4}(b) is the same as Fig.~\ref{fig:CRIB4}(a) but for $d = 1000$.

First of all, from Fig.~\ref{fig:CRIB4}, it is clear that, when $T \gamma \gtrsim 1$, fast storage efficiency, with or without CRIB, deteriorates because the spontaneous emission decreases the total efficiency by $\exp(-2\gamma T$). Optimal adiabatic storage, on the other hand, does well in this limit (provided we have a reasonable $d$). Moreover, in the adiabatic limit $T d \gamma \gg 1$, optimal adiabatic storage with homogeneous broadening is always more efficient than fast storage with or without CRIB. This follows from the fact that in the adiabatic limit ($T d \gamma \gg 1$) the error in fast storage ($\approx 1-\exp(-2 \gamma T) \approx 2 \gamma T$) is greater than the error in adiabatic storage ($\sim 5.8/d$). 

Secondly, when $T d \gamma \lesssim 1$, neither of the methods does very well. This follows directly from the general time-reversal argument described in detail in paper II. According to these arguments, the optimal storage is obtained as the time reverse of optimal retrieval. One cannot therefore store faster than one can retrieve. The fastest retrieval is obtained by using the fast retrieval method where the excitation is transferred from state $|s\rangle$ into state $|e\rangle$  with a $\pi$ pulse. With this procedure, all atoms radiate in phase and, by constructive interference, give a short output pulse of duration $T\sim 1/(d\gamma)$,  as explained in Sec.~VII of paper II. This procedure gives the fastest possible retrieval, and its time reverse is the fastest possible storage (which works optimally only for certain input mode shapes). The storage and retrieval of any mode thus becomes inefficient for $T d \gamma<1$.  The particular Gaussian mode function that we consider here does not correspond to the optimal mode and therefore its fast storage-plus-retrieval efficiency (dash-dotted line in Fig.~\ref{fig:CRIB4}) does not reach the optimal efficiency (horizontal dashed line in Fig.~\ref{fig:CRIB4}).  

Third, when $T d \gamma \lesssim 25$, reversible inhomogeneous broadening (CRIB) does help, and, with it, the fast method may do slightly better than the adiabatic method without inhomogeneous broadening. An interesting possibility is whether the controlled addition and reversing of inhomogeneous broadening could improve the adiabatic storage in the regime when the adiabaticity condition $T d \gamma \gg 1$ is not satisfied, but this investigation is beyond the scope of the present paper. 

To summarize this investigation of the possible advantages of introducing  a reversible inhomogeneous broadening, we conclude that it does provide an improvement of the efficiency, but that this improvement is limited. The fast storage technique where one just applies a resonant $\pi$ pulse without the additional broadening gives comparable results for the pulse shape that we have considered here. Moreover, even with an optimized inhomogeneous width, fast storage with CRIB performs only slightly better than optimal adiabatic storage in the original homogeneously broadened system and only when the input pulse does not satisfy the adiabaticity condition $T d \gamma \gg 1$; while, for pulses satisfying the adiabaticity condition, adiabatic storage performs better. Intuitively one might expect that a homogeneously broadened absorption line would not be able to efficiently store an input pulse with a bandwidth that is much longer than the width of this line. One could have therefore expected a large efficiency gain from the use of CRIB to shape the atomic line to match the spectral profile of the input photon wave packet. The reason why such line shaping is not necessarily much more effective than using the unmodified line is because storage is a dynamical process; therefore, the relationship between the storage capability of the medium and its continuous wave absorption spectrum is not trivial.

We should add, however, that we have here only considered a specific input shape and the picture may be different if other inputs are considered. We also emphasize that, unlike most results in this paper and in papers I and II, the efficiencies presented in this section do not represent the true optimum for our input pulse, since one could optimize, for example, the time, at which the storage $\pi$ pulse is applied. Finally, it is worth noting that if one is free to choose the shape and duration of the input pulse, then for any given $d$ and any given inhomongeous profile, as explained in Sec.~\ref{sec:InhomSol}, one can use the time-reversal iterations to find the optimal input pulse and the maximum efficiency. We have checked for a Gaussian inhomogeneous profile that at any given $d$, as the inhomogeneous width increases from zero, the maximum efficiency drops. Although we have not checked this statement for other inhomogeneous profiles, we believe it to be generally true that, if one has the freedom of optimizing the shape and duration of the input pulse, then the addition and subsequent reversal of inhomogeneous broadening only lowers the maximum efficiency.    

We note, however, that our result that the adiabatic storage efficiency is always comaparable to or better than the efficiency of fast storage (with or without CRIB) has been found assuming that in the adiabatic storage the $|g\rangle-|s\rangle$ transition is not inhomogeneously broadened. Although this holds in Doppler-broadened atomic vapors, the broadening of the $|g\rangle-|s\rangle$ transition may be hard to suppress in many of the systems considered for photon storage with CRIB (such as the rare-earth-ion-doped crystal \cite{kroll05}). Therefore, although adiabatic storage in some of these systems seems possible \cite{hemmer02}, fast storage may still be a better option. Moreover, although we showed that, for fast storage of a single input mode, adding inhomogeneous broadening may provide only small gains in efficiency, these gains might be much more significant if several time-separated modes are to be stored together and the shape of the inhomogeneous profile is allowed to be optimized. Further investigations are required to clarify these issues.

\section{Summary \label{sec:inhomsum}}

In conclusion, we have extended in this paper the analysis of photon storage in papers I and II to include the effects of inhomogeneous broadening. In particular, we showed that in Doppler-broadened atomic vapors, at high enough optical depth, all atoms contribute coherently as if the medium were homogeneously broadened. We also showed that high-efficiency photon storage (error scaling as $\sim 1/d'^2$, where $d'$ is the observed optical depth) can be achieved in solid state systems by creating a stationary spectrally well-localized inhomogeneous profile. Finally, we demonstrated that the addition of reversible inhomogeneous broadening (CRIB) to an originally homogeneously broadened line does provide an improvement of the efficiency of fast storage followed by fast retrieval, but that this improvement is limited: in particular, in the adiabatic limit $T d \gamma \gg 1$ optimal adiabatic storage outperforms fast storage with or without CRIB. These results aim at understanding the fundamtental limits for storage imposed by the optical depth of the medium. For a complete investigation of the photon storage problem in inhomogeneously broadened media, there are several other effects and experimental imperfections that should be included, such as, for example, velocity changing collisions during the processes of storage and retrieval in Doppler-broadened gases and imperfect synchronization between the input pulse and the storage control pulse. A study of the former is in progress. 

The presented optimization of the storage and retrieval processes in inhomogeneously broadened media leads to a substantial increase in the memory efficiency. We therefore expect this work to be important in improving the efficiencies in current experiments, where the optical depth is limited and where inhomogeneous broadening plays an important role, such as in Doppler-broadened atoms in warm vapor cells \cite{eisaman05} and in inhomogeneously broadened solid state samples \cite{afzelius06,kroll05}. 

\textit{Note added.} Recently, a related paper appeared \cite{gisin06}, which discusses some of the issues considered in this paper.
\section{Acknowledgments}
We thank M.~Fleischhauer, E.~Polzik, J.~H.~M\"{u}ller, M.~D.~Eisaman, I.~Novikova, D.~F.~Phillips, R.~L.~Walsworth, M.~Hohensee, M.~Klein, Y.~Xiao, N.\ Khaneja, A.~S.~Zibrov, P.~Walther, A.~Nemiroski, and M.~Afzelius for fruitful discussions. This work was supported by the NSF, Danish Natural Science Research Council, DARPA, Harvard-MIT CUA, and Sloan and Packard Foundations.


\appendix
\section{Independence of Free-Space Retrieval Efficiency from Control and Detuning \label{sec:appretr}}

We showed in Sec.~\ref{sec:dopretst} that for the case of inhomogeneous broadening that leaves the $|s\rangle-|g\rangle$ transition homogeneously broadened, the retrieval efficiency in the cavity model is independent of the control field and the detuning (provided all excitations are pumped out of the atoms). In this appendix, we show the the same result holds in the free-space model.

We consider a general situation in which $S_j(\tilde z, \tilde t=0)$ are not necessarily equal for different velocity classes $j$ (we assume here that retrieval begins at $\tilde t=0$). Laplace transforming Eqs.~(\ref{doppler2eqe})-(\ref{doppler2eqs}) in space ($\tilde z \rightarrow u$) and using $\eop_\textrm{in} = 0$, we obtain
\begin{eqnarray} 
\eop &=& i \frac{\sqrt{d}}{u} P,
\\ \label{pj}
\partial_{\tilde t} P_j &=& - (1 + i (\tilde \Delta+\tilde \Delta_j)) P_j - \frac{d}{u} \sqrt{p_j} P + i \tilde \Omega S_j,
\\ \label{sj}
\partial_{\tilde t} S_j &=& i \tilde \Omega^* P_j,
\end{eqnarray}
where $P = \sum_k \sqrt{p_k} P_k$. The retrieval efficiency is then
\begin{equation} \label{effdopfree}
\eta_\textrm{r} = \sum_{j,k} \sqrt{p_j p_k} \mathcal{L}^{-1} \left\{\frac{d}{u u'} \int_{0}^\infty d \tilde t P_j(u, \tilde t) \left[P_k(u'^*, \tilde t)\right]^*\right\},
\end{equation}
where $\mathcal{L}^{-1}$ stands for inverse Laplace transforms $u \rightarrow \tilde z$ and $u' \rightarrow \tilde z'$ with the evaluation of both at $\tilde z = \tilde z' = 1$. From Eqs.~(\ref{pj}) and (\ref{sj}), it follows that
\begin{eqnarray} \label{pjpkfree}
&\frac{d}{d  \tilde t}\left(P_j(u,  \tilde t) \left[P_k(u'^*, \tilde t)\right]^*\! + S_j(u,  \tilde t) \left[S_k(u'^*,  \tilde t)\right]^*\right) =&
\nonumber \\
&- (2+i (\Delta_j - \Delta_k)) P_j(u,\tilde t) \left[P_k(u'^*,\tilde t)\right]^*& 
\\\nonumber 
&-\frac{d}{u} \sqrt{p_j} P(u,\tilde t) \left[P_k(u'^*,\tilde t)\right]^*\!\!-\! \frac{d}{u'} \sqrt{p_k} P_j(u,\tilde t) \left[P(u'^*,\tilde t)\right]^*\!\!\!.&
\end{eqnarray}
If $M$ is the number of velocity classes, Eq.~(\ref{pjpkfree}) stands for $M^2$ equations in $M^2$ variables $P_i P_k^*$. We can write them in a matrix form and, in principle, invert the $M^2 \times M^2$ matrix on the right-hand side and thus solve for $P_j P^*_k$ as a linear combination of $\frac{d}{dt} (P_a P^*_b + S_a S^*_b)$ for various $a$ and $b$. Inserting this into Eq.~(\ref{effdopfree}), applying the fundamental theorem of calculus, and assuming the retrieval is complete (i.e., no excitations remain in the atoms), the retrieval efficiency can be expressed as $\mathcal{L}^{-1}$ of a linear combination (with $u$- and $u'$-dependent coefficients) of $S_a(u, 0) \left[S_b(u'^*, 0)\right]^*$, and is thus independent of control and detuning.


\begin{thebibliography}{99}
\bibitem{fastnote} Adiabatic storage and retrieval with a step (i.e., constant) control profile can also be solved using this technique. However, we will not pursue this investigation in the present paper.
\bibitem{nopipulse} For clarity, we note that, although fast storage is indeed mentioned in Ref.~\cite{kraus06} as a way of extending storage time beyond $1/\gamma$, the authors mainly focus on using only two levels and replacing the application of two $\pi$ pulses necessary for fast storage and retrieval with an application to each atom of a position-dependent phase shift $|g\rangle \rightarrow \exp(2 i \omega_{eg} z/c) |g\rangle$, where $z$ is the position of the atom. The treatment of this method is, however, mathematically identical to the treatment of fast storage with perfect $\pi$ pulses and no decoherence during storage. Reference \cite{kroll05}, in contrast, fully concentrates on what we call fast storage.
\end{thebibliography}
\end{document}